\documentclass[12pt]{iopart}

\newcommand{\gev}{\rm GeV}

\begin{document}

\article{}{Unified model for inflation and dark energy}
\author{Gabriel Zsembinszki}

\address{Grup de F{\'\i}sica
Te{\`o}rica and Institut de F{\'\i}sica d'Altes
Energies\\Universitat Aut{\`o}noma de Barcelona\\
08193 Bellaterra, Barcelona, Spain} \ead{gabrielz@ifae.es}

\begin{abstract}
We present a model based on the idea that Planck-scale effects may
produce a small explicit breaking of global symmetries. The model
contains a new complex scalar field $\Psi$, charged under a
certain global $U(1)$ symmetry, interacting with  a new real
scalar field $\chi$, neutral under $U(1)$. For exponentially small
breaking, the model accounts for both {\em early} and {\em late}
periods of acceleration of the universe.

\end{abstract}


\maketitle
\section{Introduction}
It is well known that symmetries that are broken at a given energy
scale may be restored at higher energies. The Standard Model (SM)
of particle physics based on the gauge group $SU(3)\times SU(2)
\times U(1)$ describes very well the physics in a wide range of
energies, but can only be tested directly up to the highest energy
scales reached in particle accelerator experiments. Because of
this limitation, we are unable to probe the physics at higher
energies in accelerators and see if the SM is still working, or
one has to go beyond it. Nevertheless, the universe itself may be
regarded as a huge "laboratory", since when it was just a small
fraction of a second old, the energy it contained was enormous,
much larger than the maximum energy that can be reached in
terrestrial experiments. These first moments of the universe left
some imprints on the present observable universe, so that,
indirectly, we are able to get some hints about the physics of
those huge energies. Some of these hints raise many deep questions
for which the SM is unable to give the right answer and this is
why many physicists are looking for extensions of the SM. As we
are trying to extend the theory from lower to higher energies, it
is natural to suppose the existence of additional symmetries,
either local, or global. There has been a lot of interest in
studying global symmetries at high energies
\cite{Hartle:1983ai,Giddings:1987cg,Coleman:1988tj,Giddings:1988cx}.
There are reasons to expect that quantum gravity effects break
global symmetries: global charges can be absorbed by black holes
which may evaporate, "virtual black holes" may form and evaporate
in the presence of a global charge, wormholes may take a global
charge away from our universe to another one, etc.

In \cite{Giddings:1987cg}, the loss of quantum coherence in a
model of gravity coupled to axions is investigated. The loss of
coherence opens the possibility that currents associated with
global symmetries are not exactly conserved, while those
associated with local symmetries are still exactly conserved.
Coleman \cite{Coleman:1988tj} argued that incoherence is not
observed in a many-universe system in an equilibrium state, and
pointed out that if wormholes exist they can explain the vanishing
of the cosmological constant. The authors of
\cite{Giddings:1988cx} pointed that even if incoherence is not
observed in the presence of wormholes, other interesting
consequences may emerge, such as the appearance of operators that
violate global symmetries, of arbitrary dimensions, induced by
baby universe interactions. In this context, the authors of
\cite{Kallosh:1995hi} argue that if global symmetries are broken
by virtual black holes or topology changing effects, they have to
be exponentially suppressed. In particular, in order to save the
axion theory, the suppression factor should have an extremely
small value $g<10^{-82}$. This suppression can be obtained in
string theory, if the string mass scale is somewhat lower than the
Planck scale, $M_{\rm str}\sim 2\times 10^{18} \gev$. Thus we
expect to have an exponential suppression of the explicit breaking
of global symmetries, but, as summarized here, even with such
extremely small breaking very interesting cosmological effects may
appear \cite{Masso:2004cv,Masso:2006yk}.

\section{The Model}

In the present contribution we would like to explain our model,
which is able to describe both {\bf early} and {\bf late}
acceleration periods of the universe. The first one is supposed to
have occurred in the very early universe and was given the name
{\bf inflation}. The need for an inflationary period of expansion
is related to various problems in cosmology, e.g., the flatness
and horizon problems, the small-scale inhomogeneities, unwanted
relics, etc. The second period of accelerated expansion has
started recently, at a redshift $z\sim \Or(1)$, and is generically
attributed to the so-called {\bf dark energy}. It is suggested by
observations of type Ia supernovas \cite{Supernova}, the matter
power spectrum of large scale structure \cite{LSS} and anisotropy
of the cosmic microwave background radiation \cite{WMAP3}.

In our model, the fields responsible for the two accelerating
phases are components of a new complex scalar field $\Psi$ that is
charged under a certain global $U(1)$ symmetry, with spontaneous
breaking scale $f$. We may write the field $\Psi$ as

\begin{equation} \Psi=\phi\, \rme^{\rmi\theta/f} \label{Psi}
\end{equation}
and identify the inflaton with the radial part $\phi$, and the
dark energy field with the angular part $\theta$. We also have a
potential containing the following $U(1)$-symmetric term

\begin{equation}
V_1(\Psi) =\frac14\lambda\left(|\Psi|^2-f^2\right)^2\label{V1}
\end{equation}
where $\lambda$ is a coupling constant.

In order to satisfy all the constraints to be imposed on any
realistic inflationary model, we also introduce a real scalar
field $\chi$, neutral under $U(1)$, which interacts with the field
$\Psi$ and assists $\phi$ to inflate. The interaction term is
$U(1)$-symmetric and has the form
\begin{equation}
V_2(\Psi,\chi)=\frac12m_{\chi}^2\chi^2+\left(\Lambda^2-
\frac{\alpha^2|\Psi|^2\chi^2}{4\Lambda^2}\right)^2\label{V2}
\end{equation}
with $\alpha$ being a coupling constant, and $\Lambda$ and
$m_{\chi}$ are some arbitrary mass scales. Until here, the
effective potential only contains $U(1)$-symmetric terms, so that
the sum of (\ref{V1}) and (\ref{V2}) represents the symmetric part
of the effective potential
\begin{equation}
V_{\rm sym}(\Psi,\chi) = V_1(\Psi) + V_2(\Psi,\chi).\label{Vsym1}
\end{equation}

We wish to study the consequences of allowing terms in the
potential, which {\it explicitly} break $U(1)$. Without knowing
the details of how Planck-scale physics breaks global symmetries,
we introduce the most simple effective $U(1)$-breaking term
\cite{Planck-scale-axion}
\begin{equation}
 V_{\rm non-sym}(\Psi)=-g\frac1{M_{\rm P}^{n-3}}|\Psi|^n\left(\Psi
\rme^{-\rmi\delta}+\Psi^{\star}
\rme^{\rmi\delta}\right)\label{Vnon_sym1}
\end{equation}
where $M_{\rm P}\simeq 1.22\times 10^{19}\, \gev$ is the Planck
mass and $n$ is an integer satisfying $n>3$. Because we expect the
coupling $g$ to be exponentially suppressed \cite{Kallosh:1995hi},
we consider (\ref{Vnon_sym1}) as a small perturbation to the
symmetric term $V_{\rm sym}$. As a consequence, the non-symmetric
term can be safely neglected when discussing inflation, but it
plays a crucial role at present, being associated with the recent
dominating dark energy of the universe.

Thus, the effective potential of our model is given by
\begin{equation}
V_{\rm eff}(\Psi,\chi) = V_{\rm sym}(\Psi,\chi) + V_{\rm
non-sym}(\Psi)+C\label{V_tot}
\end{equation}
where $C$ is a constant that sets the minimum of the effective
potential to zero.

\subsection{Inflation}
As mentioned above, when discussing inflation in our model, the
explicit $U(1)$-breaking term $V_{\rm non-sym}$ can be neglected
and we only take into discussion the symmetric part, namely
$V_{\rm sym}$, of the effective potential. After introducing
(\ref{Psi}) in the expression for $V_{\rm sym}$, we obtain the
potential only in terms of the fields $\phi$ and $\chi$
\begin{equation}
V_{\rm sym}(\phi,\chi)=\Lambda^4+\frac12\left(m_{\chi}^2
-\alpha^2\phi^2 \right) \chi^2+
\frac{\alpha^4\phi^4\chi^4}{16\Lambda^4}
+\frac14{\lambda}(\phi^2-f^2)^2. \label{Vsym2}
\end{equation}
This potential is of inverted hybrid inflation type; {\it hybrid}
because the inflaton interacts with the auxiliary field $\chi$
whose vacuum energy dominates during inflation, and {\it inverted}
because the inflaton mass term has the changed sign, as compared
to typical hybrid inflation models. The mechanism that drives
inflation in hybrid models is well described in the literature
\cite{Hybrid-Inflation}.

There are mainly three constraints that should be imposed on our
model in order to fit observations:
\begin{itemize}

\item
   sufficient number of e-folds of inflation $N(\phi)
 =(8\pi)/M_{\rm P}^2\int_{\phi_{\rm end}}^{\phi}
(V_{\rm sym}/V'_{\rm sym})\rmd\phi$ in order to solve the flatness
and the horizon problems. Here, $\phi_{\rm end}\equiv
m_{\chi}/\alpha$ is the value of the inflaton field at the end of
inflation. The number of e-folds that occur after a given scale
leaves the horizon is given by \cite{inflation}
\begin{equation}
N\simeq 62-\ln\frac{k}{a_0 H_0}-\ln\left(
\frac{10^{16}\gev}{\Lambda^{1/4}} \right)+\frac13\ln \left(
\frac{T_{\rm rh}}{\Lambda^{1/4}} \right)\label{efolds}
\end{equation}
where $k$ is the scale that exits the horizon, $T_{\rm rh}$ is the
reheating temperature and the biggest explored scale is the
present Hubble distance $a_0/k=H_0^{-1}=6000$ Mpc;

\item
   the amplitude of the primordial curvature power-spectrum
   produced by quantum fluctuations of the inflaton field should
   fit the observational data \cite{WMAP3}
   \begin{equation}
       {\cal P_R}^{1/2}=\sqrt{\frac{128\pi}3}\frac{V_{\rm sym}(\phi_0,0)^{3/2}}
       {M_{\rm P}^3 V'_{\rm sym}(\phi_0,0)}\simeq 4.86\times 10^{-5}\label{curvature}
   \end{equation}
where a prime means derivative with respect to $\phi$ and $\phi_0$
is the value of the inflaton field when the scale $k_0=0.002 {\rm
Mpc}^{-1}$ exits the horizon;

\item
   the value of the spectral index $n_{\rm s}\simeq 1-6\epsilon+2\eta$ should be
   in the allowed range suggested by the recent three-year Wilkinson Microwave Anisotropy
   Probe data \cite{WMAP3},
   $n_{\rm s}=0.951^{+0.015}_{-0.019}$. Here $\epsilon\equiv
   (M_{\rm P}^2/16\pi)(V'_{\rm sym}/V_{\rm sym})^2$ and $\eta
 \equiv (M_{\rm P}^2/8\pi)(V''_{\rm sym}/V_{\rm sym})$ are slow-roll parameters.
 In what follows, we will neglect the $\epsilon$ term in the expression of
 $n_{\rm s}$ for simplicity, which is a fairly good approximation (less than $0.1\%$
 effect on $n_{\rm s}$). We obtain
\begin{equation}
         \frac{M_{\rm P}^2}{4\pi}\frac{V''_{\rm sym}(\phi_0,0)}{V_{\rm sym}(\phi_0,0)}
         \simeq -0.05.\label{spectral_index}
   \end{equation}
  \end{itemize}
Combining (\ref{efolds}), (\ref{curvature}) and
(\ref{spectral_index}) we can obtain the dependences of $\lambda$,
$\Lambda$ and $\phi_0$ on the scale $f$. The other parameters of
our model, namely $m_{\chi}$ and $\alpha$, do not appear in the
previous equations, but there are some relations between them that
should be satisfied in order for the hybrid inflation mechanism to
work: $\alpha\gg\lambda$, $m_{\chi}<\alpha f$ and the more
important relation
\begin{equation}
f<M_{\rm P}\label{limit-sup-f}
\end{equation}
which sets an upper limit for $f$. Thus, given the value of $f$,
one is able to calculate the values of the other parameters of our
model. Finally, the parameter $g$ does not appear in the
discussion about inflation, but will be important when discussing
dark energy in our model.

\subsection{Dark energy}

After inflation, the fields $\phi$ and $\chi$ settle down at the
minimum of the symmetric part of the effective potential, the only
part "surviving" after inflation being the non-symmetric term
$V_{\rm non-sym}(\theta)= -2\,g(f/M_{\rm P})^{n-3}M_{\rm
P}^4\cos{(\theta/f)}$. The angular field $\theta$ can take any
value in the range $(0,2\pi f)$, after the end of inflation. If
the potential of $\theta$ is sufficiently flat, $\theta$ may have
a slow-rollover up to the present time, such that it may act as a
quintessence field and explain the dark energy of the universe.
For this to happen, there are two conditions to be satisfied:
\begin{itemize}
\item
  slow-rolling of $\theta$ at present, given by the condition
\begin{equation}
m_{\theta}< 3H_0\label{DE_mass_cond}
\end{equation}
where $m_{\theta}=\sqrt{2g}\,(f/M_{\rm P})^ {(n-1)/2}M_{\rm P}$ is
the mass of $\theta$ and $H_0\sim 10^{-42}\;\gev$ is the Hubble
parameter today (Hubble constant);

\item
  the energy density of $\theta$ should be comparable with the present
  critical density $\rho_{{\rm c}_0}$
\begin{equation}
\rho_{\theta}\simeq V_{\rm non-sym}(f,f)\sim \rho_{{\rm
c}_0}\equiv\frac{3H_0^2M_{\rm P}^2}{8\pi}\label{DE_energy_cond}
\end{equation}
where we supposed that both $\phi$ and $\theta$ are of $\Or(f)$
today.
\end{itemize}
Combining (\ref{DE_mass_cond}) and (\ref{DE_energy_cond}) we
finally obtain \cite{Masso:2006yk}
\begin{equation}
f > \frac16 M_{\rm P}\label{inf_limit_f}
\end{equation}
\begin{equation}
g< \frac{3\times 6^{n+1}}{8\pi}\frac{H_0^2}{M_{\rm
P}^2}.\label{sup_limit_g}
\end{equation}

\section{Discussions and Conclusions}

We have presented a model that can explain {\em both} inflation
and dark energy by introducing a new complex scalar field $\Psi$
with a potential invariant under a new global $U(1)$ symmetry. The
radial part of $\Psi$, namely $\phi$, is responsible for inflation
when assisted by a new real scalar field, $\chi$, neutral under
$U(1)$. The angular field, $\theta$, acquires a tiny mass due to a
small explicit breaking of the potential and acts as a
quintessence field, explaining the nature of the present
dominating dark energy of the universe. The explicit symmetry
breaking comes from quantum gravity effects, but we expect them to
be exponentially suppressed.

Let us see how much suppression is needed in our model. Equations
(\ref{inf_limit_f}) and (\ref{limit-sup-f}) indicate that $f$
should be close to the Planck mass. For definiteness, we set it to
$f=0.5M_{\rm P}$. With $\alpha=10^{-2}$, $m_{\chi}=3\times
10^{16}\gev$ and the lowest possible $n=4$, we obtain for the
other parameters of our model the values $\phi_0=0.135 f$,
$\lambda=9\times 10^{-14}$, $\Lambda=5.6\times 10^{15}\,\gev$ and
the limit $g<10^{-119}$. The constraint on $g$ suggests the level
of suppression of the effects of quantum gravity in breaking our
$U(1)$ global symmetry. Nevertheless, such tiny values lead to
interesting effects for cosmology, e.g., the field $\theta$ acts
like a quintessence field at present. As a final remark, we should
say that this slow-roll regime will not last forever, since when
$m_{\theta}\simeq 3H$, the slow-roll regime will end and $\theta$
will start to rapidly oscillate around its minimum.

\ack

I would like to thank Eduard Mass{\'o} for his help. This work is
supported by the Spanish grant FPA-2005-05904 and by Catalan DURSI
grants 2005SGR00916 and 2003FI00138.

\end{document}